# Purcell enhancement of a cavity-coupled emitter in hexagonal boron nitride

Johannes E. Fröch,[1,*] Chi Li,[1] Yongliang Chen,[1] Milos Toth,[1,2] Mehran Kianinia,[1,2] Sejeong Kim,[3,*] Igor Aharonovich[1,2]

[1]School of Mathematical and Physical Sciences, University of Technology Sydney, Ultimo, New South Wales 2007, Australia

[2]ARC Centre of Excellence for Transformative Meta-Optical Systems (TMOS), University of Technology Sydney, Ultimo, New South Wales 2007, Australia

[3]Department of Electrical and Electronic Engineering, University of Melbourne, Victoria, 3010, Australia

**Corresponding**
j.e.froech@gmail.com; sejeong.kim@unimelb.edu.au

**Abstract**

Integration of solid state quantum emitters into nanophotonic circuits is a critical step towards fully on-chip quantum photonic based technologies. Among potential materials platforms, quantum emitters in hexagonal boron nitride have emerged over the last years as viable candidate. While the fundamental physical properties have been intensively studied over the last years, only few works have focused on the emitter integration into photonic resonators. Yet, for a potential quantum photonic material platform, the integration with nanophotonic cavities is an important cornerstone, as it enables the deliberate tuning of the spontaneous emission and the improved readout of distinct transitions for that quantum emitter. In this work, we demonstrate the resonant tuning of an integrated monolithic hBN quantum emitter in a photonic crystal cavity through gas condensation at cryogenic temperature. We resonantly coupled the zero phonon line of the emitter to a cavity mode and demonstrate emission enhancement and lifetime reduction, with an estimation for the Purcell factor of ~ 15.

**Introduction**

Integrated quantum photonics circuitry offers unique opportunities and advanced modalities in the fields of quantum information processing, communications, and sensing.[1] This is made possible by the potential scalability of solid state systems, where in the ideal scenario the on-chip routing of single photons and the on-chip manipulation of qubits enables fast and low-loss operations.[2,3] Specifically, a distinct opportunity arises from interfacing quantum light sources with nanophotonic cavities and optoelectronic device architectures to enhance optical readout and enable on-chip manipulation with various demonstrations over the last years that have shown the disruptive potential of this approach.[4-7] Accordingly, substantial efforts have been put forth towards chip-scale integration of solid state single photon emitters (SPEs) with photonic crystal cavities (PCCs), waveguides and on-chip detectors.[8-14] A key stepping stone in this effort is the controlled alteration of the spontaneous emission by tuning a single emitter into and out of resonance with a cavity mode.[15-17]

SPEs in hexagonal boron nitride (hBN) have recently been identified as potential candidates for integrated quantum photonic applications due to their superior brightness, narrow bandwidth, ease of engineering and access to spin initialization and readout.[18-26] Whilst effort has been put forth in the integration into dielectric,[27-33] plasmonic,[34-39] or optoelectronic devices,[40,41] the critical milestone of coupling quantum emitters to monolithic hBN cavities and a controlled demonstration of a Purcell enhancement has remained elusive.[42]

Herein we demonstrate the controlled coupling of a hBN SPE to a monolithic 1D PCC. We observe a 10-fold intensity enhancement of the emitter's zero phonon line (ZPL) when it is tuned to the cavity mode, along with a lifetime reduction factor of 1.5, for which we derive a Purcell factor of 15. Our results advance the realization of on-chip integrated quantum photonics platform with hBN.

**Results**

To realize coupling of hBN SPEs to monolithic resonators, we developed an optimized nanofabrication protocol for devices from exfoliated crystals, suitable to produce large yields of high quality 1D PCCs. In comparison to our prior work the process flow is enhanced by 2 key factors, which produced smooth sidewalls in a single etch cycle and facilitate a wet-chemical undercut of the devices. The fabrication process in brief (detailed in Methods) consists of 1) exfoliation of hBN flakes (HQ Graphene) onto silicon substrates, 2) precleaning, 3) high temperature cleaning and oxidation of the silicon substrate, 4) identification of suitable flakes with a thickness of ~ 200 nm, 5) electron beam lithography (EBL), 6) reactive ion etching

(RIE), 7) wet-chemical undercut in a KOH solution, 8) post fabrication cleaning and annealing. We emphasize here the importance of steps (3) and (6) as the two factors that enabled the successful resonant coupling in this study. In detail, during step (3), a two-step annealing treatment was used, consisting of 1 hour at 650 °C and 30 min at 850 °C, both in air. During this step the adhesion of the flakes to the substrate increased substantially and the uncovered silicon surface preferentially oxidized as compared to regions beneath hBN flakes. This in turn allowed the selective wet-chemical undercut in step (7). Specifically, due to a high etch selectivity of KOH for Si relative to $SiO_2$, only regions beneath the patterned regions etched during this step, as apparent in scanning electron microscope (SEM) and optical microscope (OM) images in Fig. 1(a). Secondly, we improved the RIE conditions in step (6). Whilst previously reported etching conditions for hBN devices in RIE and ICP-RIE utilized high ICP powers and high flow rates for the chemical etch component, these conditions resulted in angled sidewalls and therefore smaller features, such as airholes in a PCC could not be sufficiently etched.[42-44] In contrast, we found that in an adverse parameter space, with lower flow rates for the chemical etch component and a higher RIE/ICP power ratio, the sidewalls could be fully etched in a single etch cycle, as evident from top and oblique angled SEM images in Figs. 1(b) and (c), respectively.

The intended design of the 1D PCC pursued in this work consists of a periodic arrangement (230 nm) of circular air holes in a 200 nm thick and 300 nm wide nanobeam, whereas airholes at the center are detuned in size and spacing following a parabolic shape to form a photonic well, as shown in Fig. 1(d). The photonic resonator yields a high field enhancement in the very center of the device, as depicted by finite-difference time-domain (FDTD) simulations in Figs. 1(e) and (f) for the cross sectional views through the x-z and x-y plane, respectively. Notably, by employing a monolithic approach, an SPE could be placed at the center of the device as compared to hybrid approaches, thus facilitating higher degree of light-matter coupling. As typical device characteristics from simulation, we derived a quality factor (Q-factor) of ~ $5 \times 10^5$ and a mode volume of ~ 1 $(\lambda/n)^3$. However, as typically observed, the experimental Q-factor is naturally limited by fabrication and material constraints, i.e., sidewall roughness, deviations from the intended design, and material absorption. Nevertheless, as shown in Fig. 1(c), thanks to the improved fabrication conditions, we observed quality factors up to ~ 5,000 (Fig. 1(g)), which are higher as compared to the best values from previous works with monolithic hbn photonic crystal cavities (~ 2,000)[42, 43] or other types of hBN resonators[45, 46].

We emphasize that the above-described development for an improved fabrication protocol is relevant beyond the scope of this work to systems studying hBN for their phonon polariton properties,[47] nonlinear response,[48, 49] as well as a platform for enhanced integrated 2D materials coupling.[45, 46, 50] In particular, with future development of wafer scale growth of high quality hBN,[51, 52] extended photonic architectures will become feasible, whilst limiting factors such as grain boundaries in exfoliated flakes will be eliminated.

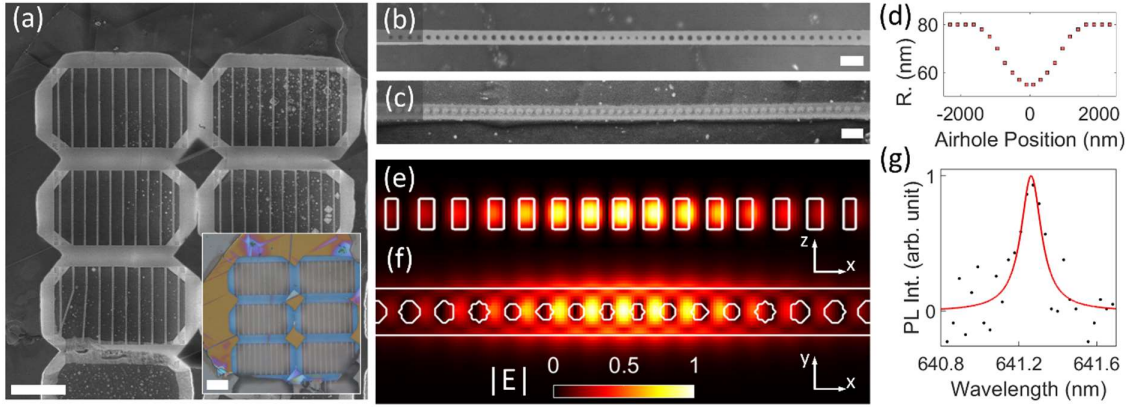

*Figure 1. Optimized monolithic 1D Photonic Crystal Cavities from hBN. (a) Representative SEM image of a cavity array. The inset shows the optical microscope image of the same flake. The optical contrast of the hBN flake relative to the Si substrate is clearly discernible. Scale bars in both images correspond to 10 μm. (b) and (c) show high resolution SEM images of a single freestanding photonic crystal cavity from top and side view. Scale bars corresponds to 500 nm. (d) Detuning of the airhole radius (R.) as function of the position along the cavity. The center of the device was set at 0. (e) and (f) display the simulated normalized field intensity through the device center in the x-z and x-y plane, respectively. (g) Spectrum and Lorentzian fit of the device with the highest Q-factor measured with a value of ~ 5,000.*

In the following we identified suitable SPEs at the center of fabricated devices in a lab-built confocal microscope setup. Specifically, samples were studied at low temperature (~80 K) in high vacuum (~$10^{-6}$ mBar) using a cw-laser (532 nm) as the photoluminescence (PL) excitation source. We note that the coupling study was carried out at a readily accessible temperature of 80 K using liquid nitrogen cooling. This is a unique feature of the hBN system, owing to its narrow lines and high brightness. On the other hand, the ultrafast spectral diffusion at these temperatures is limited by the emission linewidth, which in turn enables proper observation of the Purcell enhancement.

Whilst we identified several defects in the photonic devices, we will focus on one emitter-cavity system where the cavity mode was in suitable spectral vicinity of the ZPL and spatially located at the center of the device and a clear intensity enhancement was observable upon tuning. As shown in Fig. 2(a) the in-detail studied emitter (denoted as SPE) displayed a ZPL at ~ 610 nm (Full width at half maximum ~ 3 nm). Further in the spectrum we observed other emitters, with ZPLs at ~ 600 nm, and 580 nm. Narrower peaks at ~ 574 nm, 598 nm, 618 nm, and 620 nm correspond to the hBN Raman line, as well as fundamental ($C_0$) and higher order resonances ($C_1$, $C_2$) of the PCC, respectively. The fundamental cavity mode at 597 nm displayed a Q-factor of ~2,500 (inset Fig. 2(a)). Important for our experiment, the SPE at ~ 610 nm did not display any broad spectral diffusion or blinking within the time frames of the experiment, as shown in a spectral series in Fig. 2(b). Intensity variations in the spectral series correspond to slow mechanical drifts occurring in our custom PL system, whereas the drift was compensated by adjusting the position of a scanning mirror whilst keeping count rate constant. Furthermore, the intensity variation in the spectral series displays no correlation between other emitters at ~ 580 nm and the studied SPE, thus excluding the possibility that the defects are related.

The quantum nature of the emitter was confirmed by spectrally filtering the PL emission around the ZPL ((610 +/- 5) nm) and analyzing the second-order autocorrelation curve ($g^{(2)}(t)$) using a Hanbury-Brown and Twiss setup. At zero delay time, the measurement (Fig. 2(c)) reveals a value of ~ 0.3, which is higher than 0 likely due to the presence of other nearby luminescent defects, particularly a defect with a broader ZPL at ~ 600 nm.

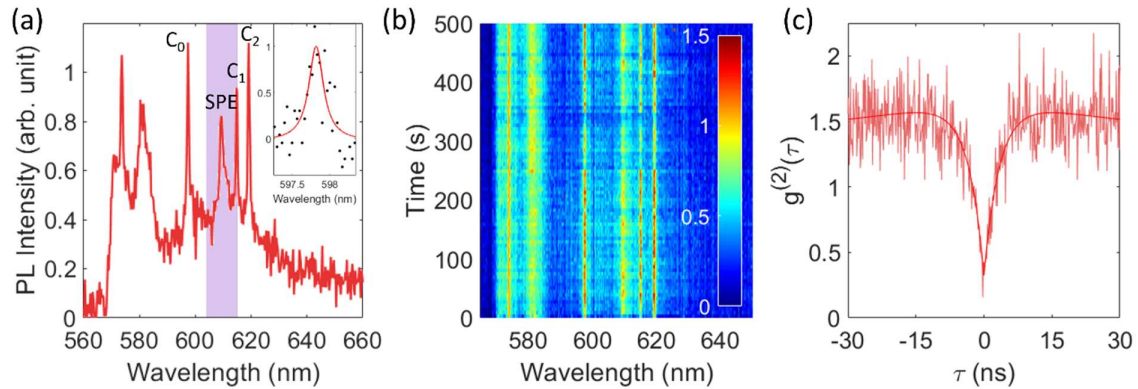

*Figure 2. Pre-characterization of the SPE before tuning. (a) Spectrum displaying the identified SPE at 610 nm in a hBN PCC with a fundamental resonance ($C_0$) at ~ 598 nm (inset) and further resonances $C_1$, $C_2$ at ~ 618nm, and 620 nm, respectively. The shaded region displays the spectral range that was selected for autocorrelation and lifetime measurements.*

*(b) Spectral series normalized to the Raman line at 574 nm. (c) Second-order autocorrelation measurement of the emitter in the spectral window indicated in (a).*

To couple the SPE to the cavity resonance, we utilized gas condensation tuning of the mode.[53] For this purpose, ambient air was delivered to the substrate by a capillary, which was oriented towards the sample surface. As the gas condensed on the surface the effective size of the resonator increased and the mode red-shifted sequentially over several tuning steps, as shown as a spectral series in Fig. 3(a). In total, we achieved a tuning range of ~ 10 nm for the cavity mode.

Notably, as the cavity mode and the SPE were tuned into resonance we observed a ~ 10-fold enhancement of the ZPL emission compared to the off-resonant emitter (red spectrum, averaged over 5 steps), as shown in Fig. 3(b). Further insight into the coupled system was obtained by fluorescence lifetime measurements using a picosecond pulsed laser, presented in Fig. 3(c) as red (blue) circles for the off (on) resonant emissions and fitted with a single exponential curve. Notably, both decay curves clearly exhibit two components, whilst in the coupled state the faster decay component reduced significantly. Conversely, the slower decay did not change substantially upon coupling and is therefore attributed to background luminescence. From a single exponential fit to the faster component we deduced an intrinsic fluorescence lifetime of ~ 1 ns for the emitter, which reduced to 0.67 ns, as it was coupled to the resonance.

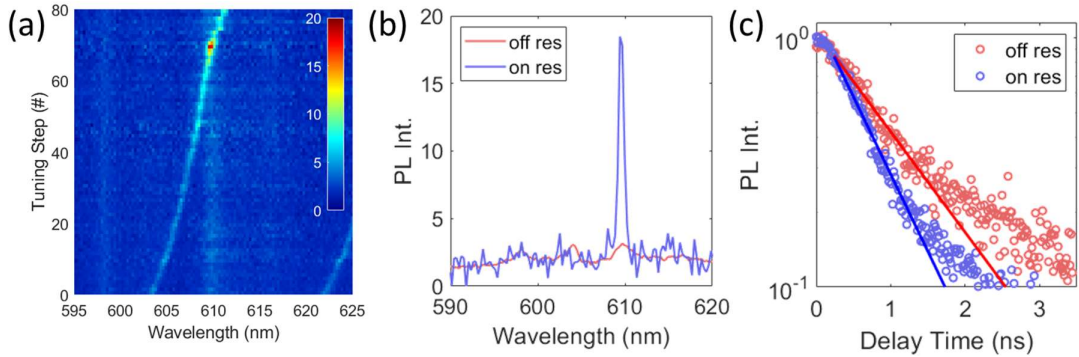

*Figure 3. Resonant coupling of an hBN SPE. (a) Kinetic spectral series of the cavity mode (starting at ~ 600 nm) tuned into resonance with a ZPL of an SPE at 610 nm. (b) Direct comparison of PL spectra, for the emitter off resonance (average over 5 spectra, red curve) and on resonance (blue curve), yielding a PL enhancement of 10. (c) Fluorescence lifetime measurement of the emitter off (on) resonance, shown by the red (blue) curves. The data was fitted with a single exponential decay model.*

To derive a Purcell factor for the coupled decay, we consider that the linewidth of the emitter (~ 3 nm) is significantly broader than the cavity mode (~ 0.3 nm), and thus the 1.5 lifetime reduction of the coupled system is a convolution of decay paths through the cavity resonance, non-resonant decay, and non-radiative decays. For this case, a suitable simple expression for the enhancement of the spontaneous emission rate at the cavity resonance was given by Englund et. al.[54]

$$F_P = \frac{I_C(\lambda)\tau_0}{I_0(\lambda)\tau_C} \quad (1)$$

Here $F_P$ denotes the Purcell factor, $I_C(\lambda)$ and $\tau_C$ are the PL intensity and lifetime on resonance, whilst $I_0(\lambda)$ and $\tau_0$ are the off-resonance intensity and lifetime, respectively. We thus obtain a Purcell factor of ~ 15, which is the highest value reported to date for a dielectric cavity-coupled hBN single photon emitter. This value is lower than the theoretically maximal value, which can be calculated by the commonly given expression for the Purcell factor (Equation (2)).

$$F_P = \frac{3}{4\pi^2}\left(\frac{\lambda}{n}\right)^3 \frac{Q}{V} \quad (2)$$

Given a Q-factor of 2,500 and a mode volume of ~ 1, at maximum one may expect a Purcell factor of ~ 190. However, this expression assumes an ideally placed quantum emitter at the center with a matching dipole to the polarization of the cavity mode at that position. Naturally, small deviations in the emitter position and a misalignment of the dipole orientation relative to the cavity resonance thus reduce the experimentally observable Purcell Factor.

**Conclusion**

In summary, we demonstrated the resonant coupling of an hBN SPE to a monolithic 1D photonic crystal cavity. We achieved this long outstanding goal, by developing an optimized nanofabrication protocol for monolithic hBN devices, primarily enabled by improved RIE parameters. The resultant structural improvements yielded higher device Q-factors as previously reported for monolithic hBN devices up to ~ 5,000. By gas condensation tuning we then demonstrated the controlled coupling of the ZPL of a hBN SPE to the fundamental cavity mode, yielding a ~ 10-fold PL enhancement and 1.5 lifetime reduction. Given that the system is limited by the emitter linewidth and non-radiative decays, we derived a Purcell factor of ~ 15. We emphasize that this value is on par with values that were typically reported for other material systems in their respective early development stage,[15] thus putting hBN on the right path for the development as a potential quantum nanophotonic material platform. Although the immediate demonstration of the SPE device coupling was performed using photonic crystal cavities fabricated from exfoliated flakes, we envision that with further process of wafer scale single crystal growth, larger integration scales will become more accessible. Overall, the results present an important milestone for the realization of integrated hBN on-chip quantum nanophotonic devices.

**Methods**

**Device Fabrication.** hBN flakes were transferred onto precleaned silicon substrates using tape exfoliation from a larger bulk crystal, purchased from HQ Graphene. After exfoliation, the tape residues were removed by calcination in air on a hot plate at 500 C for ~ 4 hours. Then the sample was annealed in a quartz tube under ambient gas. The sample was annealed for 1 hour at 650 C, and subsequently at 850 C for 30 min. After annealing suitable flakes of hBN were identified by optical contrast and confirmed by atomic force microscopy. Samples were then spin coated with a polymer resist (CSAR, AllResist GmbH.) with a thickness of ~ 500nm. 1D PCC designs were patterned into the resist using electron beam lithography at 30 keV in a Zeiss Supra with an attached RAITH EBL system. After patterning, patterns were developed and transferred into the underlying hBN and Silicon substrate using reactive ion etching in a Trion ICP-RIE system at 1mT, 10 SCCM Ar, 1 SCCM $SF_6$, 100 W RIE, 50 W ICP. Finally, the 1D PCCs were undercut wet-chemically in KOH (~ 10 % / ~ 40 C). After undercut the remaining resist was removed and the sample were annealed in a tube furnace in air for 30 min at 800 C.

**FDTD simulations**. Numerical modeling was performed by using the finite-difference time-domain (FDTD) method in Lumerical software (Lumerical Inc). 1D photonic crystal cavity is made of hBN and consist of 50 air holes. Both radius of the air holes and the distance between the air holes are modulated to create a cavity at the center of the nanobeam. The center of the nanobeam is a hBN and we defined the gap size as a distance between two center air holes. Q-factor is simulated by varying the gap size reaching theoretical Q-factor up to $5 \times 10^5$.

**Cryogenic PL measurements.** Samples were loaded into a modified cryostat system (Janis) with a 0.9 NA objective placed inside the vacuum chamber. The objective and sample were optically accessible through a Quartz window in the chamber wall. After loading the sample inside the chamber, the system was pumped to a pressure of ~ $10^{-6}$ mbar for a duration of ~ 30 hours, before the sample was cooled down to ~ 80 K by flowing liquid nitrogen through the cryostat. Scanning of the sample and laser beam steering was achieved by a scanning mirror

(Newport FSM300) and a 4f system, placed in the path between the scanning mirror and the entrance aperture of the objective. The excitation laser (532 nm) was separated from the PL by a dichroic mirror (Semrock) placed in the optical path. The collected PL was guided into a fiber that was connected either to a spectrometer (Andor) or two APDs (Excelitas), which were connected to a time correlator (PicoHarp 300/ PicoQuant) enabling autocorrelation measurements or time correlations measurements. For lifetime measurements a picosecond laser (Pilas PiL051X/ advanced laser diode systems) with an excitation wavelength of 512 nm and pulse frequency of 60 MHz was used.

## Acknowledgement
We acknowledge the Australian Research Council (CE200100010, DP190101058) and the Asian Office of Aerospace Research and Development (FA2386-20-1-4014).